\documentclass[12pt]{article}
\usepackage{pudlatex}
\usepackage{fullpage}
\usepackage{graphicx}

\newcommand\blog{\mathop{\rm blog}\nolimits}

\title{Extractors for small zero-fixing sources}

\author{Pavel Pudl\'ak\thanks{partiallly supported by grant EXPRO 19-27871X of the Czech Grant Agency} \\
Institute of Mathematics, CAS\\ Prague, Czech Republic
\and 
Vojt\v{e}ch R\"odl\thanks{partially supported by NSF
grant DMS 1764385}\\
Emory University\\ Atlanta, USA}

\begin{document}

\maketitle

\begin{abstract}

A random variable $X$ is an $(n,k)$-zero-fixing source if for some subset $V\sub[n]$, $X$ is the uniform distribution on the strings $\{0,1\}^n$ that are zero on every coordinate outside of $V$. An $\epsilon$-extractor for $(n,k)$-zero-fixing sources is a mapping $F:\{0,1\}^n\to\{0,1\}^m$, for some $m$, such that $F(X)$ is $\epsilon$-close in statistical distance to the uniform distribution on $\{0,1\}^m$ for every $(n,k)$-zero-fixing source $X$. Zero-fixing sources were introduced by Cohen and Shinkar in~\cite{cohen-shinkar} in connection with the previously studied extractors for bit-fixing sources. They constructed, for every $\mu>0$, an efficiently computable extractor that extracts a positive fraction of entropy, i.e., $\Omega(k)$ bits, from $(n,k)$-zero-fixing sources where $k\geq(\log\log n)^{2+\mu}$. 

In this paper we present two different constructions of extractors for zero-fixing sources that are able to extract a positive fraction of entropy for $k$ essentially smaller than $\log\log n$. The first extractor works for $k\geq C\log\log\log n$, for some constant $C$. The second extractor extracts a positive fraction of entropy for $k\geq \log^{(i)}n$ for any fixed $i\in \N$, where $\log^{(i)}$ denotes $i$-times iterated logarithm. The fraction of extracted entropy decreases with $i$. The first extractor is a function computable in polynomial time in~$n$ (for $\epsilon=o(1)$, but not too small); the second one is computable in polynomial time when $k\leq\alpha\log\log n/\log\log\log n$, where $\alpha$ is a positive constant.

The subject studied in this paper is closely related to Ramsey theory. We use methods developed in Ramsey theory and our results can also be interpreted as a contribution to this field.
\end{abstract}

\section{Introduction}

The theory of randomness extractors, a field in theoretical computer science, and Ramsey theory, a classical field in finite combinatorics, are closely connected. The best example of such a connection is the polynomial time construction of graphs that get very close to the bound on Ramsey numbers for graphs~\cite{BKSSW,BRSW}, the bound for which we do not have a proof based on explicit constructions. The construction which was developed in~\cite{BKSSW,BRSW} was based on concepts used in the theory of extractors. In this paper we will use methods of Ramsey theory to answer a question about certain type of extractors. Our results can be translated into the language of Ramsey theory and thus we also get new results in Ramsey theory. In this introduction, however, we will describe our results only in terms of the concepts used the theory of extractors and defer describing the Ramsey theoretical interpretation to Section~\ref{s-ramsey} after we introduce basic concepts.

A randomness extractor is, roughly speaking, a function $F$ that maps $n$ bits to $l$ bits, where $l\ll n$ in such a way that for every distribution $X$ from some class of distributions on $n$-bit strings, the output $F(X)$ is close to the uniform distribution on $l$-bit strings. A necessary condition for the existence of an extractor is that the entropy of the sources is $\geq l-o(l)$. If the only condition on the sources of randomness is a lower bound on their entropies, then $F$ needs a few additional truly random bits, called a random seed, as a part of input. There are many interesting classes of sources for which no additional random bits are needed for their extractors; such extractors are called \emph{deterministic} (in order to distinguish them from those that do need random seeds, which are called \emph{seeded extractors}). Examples of sources for which deterministic extractors have been constructed are sources that consist of two, or several, independent parts, affine sources, which are uniform distributions on affine subspaces of $\F_2^n$ of a given dimension, bit-fixing sources where all bits are fixed except of bits on some subset $V\sub[n]$, $|V|=k$, where the bits are truly random (these are special cases affine sources of dimension $k$), and zero-fixing sources, which are a special case of bit-fixing sources where all fixed bits are zeros. (For a precise definition of the latter two concepts, see the next section.)


Bit-fixing sources were introduced in the 1980s, see~\cite{Vaz85,BBR85,CGH85}. Initially the study of these sourced was connected with applications in cryptography, communication complexity and fault-tolerant computations. Recently more applications were found, in particular, in proving lower bounds on the formula size and designing compression algorithms.

In~\cite{kamp-zuckerman} Kamp and Zuckerman proved that for every $n$ and $k\leq n$ there exists an extractor that extracts $(\frac 12-o(1))\log k$ bits of entropy. As Cohen and Shinkar observed in~\cite{cohen-shinkar}, in general one cannot get more bit of entropy, because the Ramsey Theorem implies that if $n$ is sufficiently large w.r.t. $k$, then for any coloring of subsets of size at most $k$ there exists a subset $V$, $|V|=k$, such that for all $l\leq k$ the color of all $l$-subsets is the same.
  
The fist construction of an extractor for $(n,k)$-bit-fixing source with $k=o(n)$ that outputs $k^{\Omega(1)}$ bits is due to Kamp-Zuckerman~\cite{kamp-zuckerman}. This was improved to $k=\log^cn$, for some~$c$,  by Gabizon, Raz and Shaltiel~\cite{GRS06}. Their extractor also outputs almost all entropy bits $(1-o(1))k$. More recently, Cohen and Shinkar found a construction for $k=(1+o(1))\log\log n$ with $k-O(1)$ output bits, however their construction only gives functions computable in quasipolynomial time~\cite{cohen-shinkar}.

In the same paper, Cohen and Shinkar proposed to study zero-fixing extractors. Their motivation was twofold. First, impossibility results for the existence of zero-fixing extractors are also impossibility results for bit-fixing extractors. Second, constructing zero-fixing extractors seems to be an easier task, which may eventually help us to construct extractors for bit-fixing sources. And, indeed, they were able to find a polynomial time construction of an extractor for $(n,k)$-zero-fixing sources with $k=(\log\log n)^{O(1)}$ and $\Omega(k)$ output bits (i.e., they gave a polynomial time construction in the regime where only quasipolynimal time constructions are known for bit-fixing sources). Moreover, extractors for zero-fixing sources are very related to problems studied in Ramsey theory.

Our aim in this paper is to go beyond the state of art given by $k\approx\log\log n$ with extractors for zero-fixing sources. 
We will present two polynomial time constructions of extractors that produce $\Omega(k)$ bits for zero-fixing sources with where $k$ can be essentially smaller than $\log\log n$. It should be noted that for $k=o(\log\log n)$ a random function is not an extractor. So prior to our work even the mere existence of such extractors had not been known. 

Our first construction, presented in Section~\ref{s-stepping}, is based on a method of Erd\H{o}s and Hajnal~\cite{erdos-hajnal} which they used to prove lower bounds on certain infinite Ramsey numbers and which was later used to prove lower bounds on finite Ramsey numbers, see~\cite{graham-rothschild-spencer}. The basic idea is to project subsets of a set $A$ to subsets of an exponentially smaller set $B$ as follows. Assume, w.l.o.g., that the cardinality of $A$ is a power of two. Take a complete binary tree $T$ with $A$ as the set of its leaves. Let $B:=\{0,1\dts\log|A|\}$ and view it as the set of levels of $T$. Given a subset $V\sub A$, generate a subtree $T_V$ of $T$ with the set of leaves equal to $V$ and define the projection of $V$ to be indices of levels on which the inner vertices of $T$ occur. In general the projection of $V$ is not one-to-one, so one has to do something more complicated. We use this projection to reduce the construction of an extractor for zero-fixing sources to an extractor for bit-fixing sources on an exponentially smaller set. Since a construction of for doubly logarithmic bit-fixing sources is known, we obtain a polynomial time construction of extractors for triply logarithmic zero-fixing sources. 

Our second construction, presented in Section~\ref{s-shift}, is based on \emph{shift graphs,} which are certain graphs defined on $l$-tuples of elements of a set. They were also first studied on infinite sets by Erd\H{o}s, Hajnal and Rado~\cite{erdos-hajnal-rado}. These graphs have some remarkable properties, one of which is their low chromatic numbers; moreover the colorings with small number of colors can be explicitly constructed. We use these colorings to define the first stage of our extractor which condenses a positive fraction of the entropy to a set of size $\delta k$, for some $\delta>0$. The resulting distribution is very much like a bit-fixing source, so we can apply a random function to obtain a distribution close to the uniform. To find such a function requires a brute-force search, but if $k$ is small enough, it can be done in polynomial time. Furthermore, we believe that some explicit constructions of extractors for bit-fixing sources could be adapted to this end. 

These two constructions together with the previous ones for smaller $k$ show that there are polynomial time computable extractors for the whole range $\{k\ |\ \exists i\in\N.k\geq\log^{(i)}n\}$. For each fixed $i$, if $k\geq\log^{(i)}n$, the extractors produce $\Omega(k)$ bits, but with $i$ increasing, the fraction of extracted bits decreases exponentially.


Finally, in Section~\ref{s-upper-bounds} we prove an upper bound on the amount of entropy that can be extracted from small zero-fixing sources. According to this bound, if $i\leq (1-o(1))k$ then a loss of approximately $i-1$ bits of entropy is inevitable if $k\leq\log^{(i)}n$. Instead of using the Ramsey theorem as a black box, we use its \emph{proof} streamlined for our purpose. Thus we get, in particular, a better bound on the relation of $n$ and $k$ for which only $(\frac 12-o(1))\log k$ bits can be extracted than the bound proved  in~\cite{cohen-shinkar}. That said, the upper and lower bounds are still very far apart. In fact, even in the case of $k$ being the triply iterated logarithm there is a huge gap: the upper bound gives approximately $k-2$, while our constructions only give $\epsilon k$ for a fairly small $\epsilon>0$.

\section{Notation and definitions}\label{s-notation}

We will mostly use standard notation. For a positive integer $n$, $[n]$ denotes the set $\{1\dts n\}$. For a set $V$ and a positive integer $k$, ${V\choose k}$ (respectively ${V\choose \leq k}$) denotes the set of all subsets of $V$ of cardinality $k$ (at most $k$), and ${\cal P}(V)$ denotes the power set of $V$. For sets $X,Y\sub\N$, $X<Y$ means that $\max X<\min Y$.
We say that $\sigma\in\{0,1,*\}^n$ is a \emph{partial vector,} or a \emph{restriction}, and $\rho\in\{0,1,*\}^n$ is its extension, if $\rho_i=\sigma_i$ for every $i$ such that $\sigma_i\neq *$. Here, $\rho$ may be a \emph{total vector,} i.e., a vector without any $*$s.
%
%
We denote by 
\[
\exp^i_r(x):= r^{r^{\cdot^{\cdot^{\cdot^{r^x}}}}}\quad\mbox{-- tower of $i$ $r$s},
\]
the iterated exponential. We will omit $i$ if it is equal to 1. 
All logarithms in this paper are in base 2. We denote by $\log^{(i)}x$ the $i$-times iterated logarithm, and $\log^*x$ stands for the least $i$ such that $\log^{(i)}x\leq 1$.
The \emph{entropy} of a random variable $X:\Omega\to R$ is defined by
\[
{\rm H}[X]:=\sum_{r\in R}\prob[X=r]\log\frac 1{\prob[X=r]}.
\]
Note that
$
{\rm H}[X]\leq\log |R|,
$
with equality iff the values of $X$ are uniformly independently distributed. 
%
The \emph{total variation distance of probability measures}  $\mu$ and $\nu$, often called the \emph{statistical distance}, is defined by
\[ 
{\rm d}(\mu,\nu):=\sfrac 12\|\mu-\nu\|_1=\sfrac 12\sum_x|\mu(x)-\nu(x)|.
\]
Let $\mu_X$ denote the probability distribution on $R$ defined by $\mu_X(r):=\prob[X=r]$, where $R$ is the range of $X$, and let $U_R$ denote the uniform distribution on $R$. An important parameter in the theory of extractors is the distance of the probability distribution generated by a random variable $X$ from the uniform distribution on the range of $X$:
\[
{\rm d}(\mu_X,U_R)=\|\mu_X-U_R\|_1=\sfrac 12\sum_{r\in R}|\prob[X=r]-|R|^{-1}|.
\] 
If the statistical distance $d(\mu_X,U_R)$ is small, then $X$ has large entropy: for every $\epsilon>0$ there exists $\delta>0$ such that
\[
{\rm d}(\mu_X,U_R)\leq\delta\ \Rightarrow\ {\rm H}[X]\geq (1-\epsilon)\log|R|.
\]
Note that this also implies that there must be at least $|R|^{1-\epsilon}$ elements in the range of $X$.
The opposite is not true; in order to get a good upper bound on ${\rm d}(\mu_X,U_R)$, we must know that the entropy is \emph{very close} to the maximum, which is $\log|R|$.

\subsection{Sources of randomness and extractors}

In this paper, a source of randomness is either a random variable $X$, or the probability distribution $\mu_X$ associated with it. It is convenient to keep both interpretations, because random variables can be composed with functions, whereas probability distributions can be handled as vectors in $\R^R$. In this section we will view a source as a probability distribution $\mu$ on some set $R$. We imagine that it has small entropy relative to the size (cardinality) of $R$. An extractor is a function $F$ that maps $R$ to a smaller set $S$ so  that $\mu$ is mapped to a probability distribution $\nu$ where $\nu$ keeps a  substantial part of the entropy of $\mu$ and is close to the uniform distribution on $S$.

We will now explain extractors in more detail. Let $F:R\to S$. We define a mapping $L^F:\R^R\to\R^S$ as follows. Let $\alpha\in\R^R$, and for $r\in R$, let $\alpha(r)$ denote its $r$-th coordinate. Then for $s\in S$, the $s$-th coordinate of $L^F(\alpha)$ is defined by
\[
L^F(\alpha)(s)=\sum_{r;F(r)=s}\alpha(r).
\]
We note some basic properties of the function $L^F$.
\ben
\item $L^F$ is linear;
\item $L^F$ maps a probability distribution to a probability distribution: \[L^F(\mu_X)=\mu_{F(X)};\]
\item $L^F$ is contracting w.r.t. to $\ell_1$ norm, i.e., $\|L^F(\alpha)\|_1\leq\|\alpha\|_1$;
\item it follows that ${\rm d}(L^F(\alpha),L^F(\beta))\leq{\rm d}(\alpha,\beta)$.
\een

Here we confine ourselves to \emph{deterministic extractors}, which means that $F$ is a function without any additional random seed. Such extractors exist only for restricted classes of sources, sources with some particular structure. Before going into details, we suggest the reader to imagine the task of constructing an extractor as a game. In this game we know that there is randomness in the source, but we do not know where exactly. E.g., in the case of bit-fixing sources, we know that there is a subset $V$ of bits with perfect randomness, but we do not know $V$. We should prepare a function $F$ that will work, i.e., produce random bits, whatever source an enemy chooses; in the case of bit-fixing sources, this means whatever set $V$ the enemy picks.
\bdf
Let $\{X_j\}_{j\in J}$ be a family of sources with range $R$, i.e., $X_j:\Omega_j\to R$ for some $\Omega_j$, $j\in J$. We say that an $F:R\to S$ is an \emph{$\epsilon$-extractor for} $\{X_j\}_{j\in J}$ if 
\[
\rm d(\mu_{F(X_j)},U_S)   =\rm d(L^F(\mu_{X_j}),U_S)
\leq\epsilon
\] 
for every $j\in J$.
\edf
A necessary condition for the existence of an $o(1)$-extractor is that $\log|S|\leq \min_j{\rm H}[X_j]+o(1)$; in the interesting cases it is always $\log|S|< \min_j{\rm H}[X_j]$. 
In most cases that appeared in the literature the sets $R$ and $S$ are sets of all 0-1 strings of some length. The next important and well-known fact follows easily from the properties of $L^F$ listed above.
\bll{l-conv}
If $F$ is an {$\epsilon$-extractor for} $\{X_j\}_{j\in J}$, then $F$ is also an $\epsilon$-extractor for every convex combination of the sources $\{X_j\}_{j\in J}$.
\el
Note that if $X$ is a convex combination of $X_j:\Omega_j\to R$, then $\mu_X$ is a convex combination of $\mu_{X_j}$, $j\in J$, as vectors in $\R^R$. What we will need in our proofs is a slightly more general principle than Lemma~\ref{l-conv}, which also follows easily from basic principles:
\bl
Let $F$ be an {$\epsilon$-extractor for} $\{X_j\}_{j\in J}$ and let $Y$ be an arbitrary source. Let $Z$ be a convex combination of sources $X_j$ and $Y$ in which $Y$ has weight $\leq\delta$. Then $F$ is an $(\epsilon+\delta)$-extractor for $Z$. 
\el

In this paper we will construct extractors for \emph{zero-fixing sources}, but we will also need a more general class of \emph{bit-fixing sources} as building blocks.
\bdf
~
\ben
\item A random variable $X$ is an \emph{$(n,k)$-zero-fixing source} if for some vector $\sigma\in\{0,*\}^n$ with exactly $k$ stars, $X$ is the uniform distribution on vectors $s\in\{0,1\}^n$ that extend $\sigma$. Equivalently, $X$ is a uniform distribution on ${\cal P}(V)$ for some $V\sub[n]$, $|V|=k$.
\item A random variable $X$ is an \emph{$(n,k)$-bit-fixing source} if for some vector $\sigma\in\{0,1,*\}^n$ with exactly $k$ stars, and $X$ is the uniform distribution on vectors $s\in\{0,1\}^n$ that extend $\sigma$.
\een
\edf

\bl
If $F$ is an $\epsilon$-extractor for $(n,k)$-bit-fixing sources, then $F$ is also an $\epsilon$-extractor for $(n,k')$-bit-fixing sources for every $k'\geq k$.
\el
\bprf
Given $\sigma$ with $k'$ stars defining a $(n,k')$-bit-fixing source with $k'>k$, we can represent it as convex combination of $(n,k)$-bit-fixing sources by fixing some subset of $k'-k$ stars in all $2^{k'-k}$ ways.
\eprf

\section{Extractors for zero-fixing sources and Ramsey theory}
\label{s-ramsey}

In Ramsey theory mappings of the form $F:{[N]\choose l}\to [M]$  for numbers $N>2$, $l\geq 2$, and $M\geq 2$ are studied. Such a mapping is called a \emph{coloring of $l$-tuples by $M$ colors.} A subset $V\sub [N]$, $|V|>l$ is called \emph{monochromatic} if $F$ is constant on ${V\choose l}$. The finite Ramsey theorem states that for every $l\geq 2$, $M\geq 2$, $k>l$ there exists $N$ such that for every coloring $F:{[N]\choose l}\to [M]$, there exists a monochromatic set $V$ of cardinality $k$. The least such $N$ is called the \emph{Ramsey number} $R^l_M(k)$. Exact values of Ramsey numbers are only known for a few instances of small numbers, but there are good bounds on the asymptotic behavior of  $R^l_M(k)$ as a function of $k$ for fixed $l,M$. In particular, if $M=4$, then  $R^l_M(k)$ grows like $l$-times iterated exponential function. In the construction presented in the following section we will use the method by which lower bounds on Ramsey numbers were proved, the \emph{stepping-up lemma}.

We will now explain the connection with extractors. What is called a \emph{coloring} corresponds to an \emph{extractor} and \emph{subsets $V\sub[N]$} correspond to \emph{sources}. Suppose that $N< R^l_M(k)$. Then there exists a coloring $F:{[N]\choose l}\to [M]$ such that no subset $V\sub[N]$ of size $k$ is monochromatic. This can be equivalently stated as follows: for every $V$ of size $k$, if we consider $F(X)$ for a random $X\in{V\choose l}$, then this random variable $F(X)$ has nonzero entropy, i.e., we can ``extract some entropy'' (though it may be very little). 

It is also well-known that for $l=2$ and $M=2$ (but also for $M$ larger than $2$), if $N\ll R^2_2(k)$, then there exists a coloring $F$ such that for every $V\sub[N]$ of size $k$ the number of pairs of color~1 is almost the same as the number pairs of color~2. This means that one can extract almost 1 bit of entropy. One can prove a similar fact also for $M$ larger, which means that we can extract almost $\log M$ bits. We will call colorings in which colors appear with almost the same frequency \emph{balanced}. A precised definition of ``balanced'' follows.

\bdf
We say that a coloring $F:A\to[M]$ is \emph{$\epsilon$-balanced} if the distribution on $[M]$ generated by $F$ is $\epsilon$-close to the uniform distribution on $M$. In plain words this means
\[
\sum_{i=1}^M \left|\frac{|\{a\in A\ |\ F(a)=i\}|}{|A|}-\frac 1M\right|\leq\epsilon.
\]
\edf

For $l>2$ constant, very little is known about the existence of colorings $F:{[N]\choose l}\to [M]$ such that the restriction of $F$ to ${V\choose l}$ is $\epsilon$-balanced for every $V$, $|V|=k$. Our results in this paper give constructions of $\epsilon$-balanced colorings at least in some special range of parameters. We can show that for every fixed $l$, if $N\leq\exp^l(k)$, then there exists a coloring $F:{[N]\choose k}\to [M]$ such that for every $V\sub[N]$, $|V|=2k$, $F$ restricted to ${V\choose k}$ is $\epsilon$-balanced. Here we can take $M$ exponentially large in $k$ and $\epsilon$ exponentially small in $k$. The precise statement follows.

\begin{corollary}[of Theorem \ref{t-shift-extractor}]
For every $l\in\N$, there exists $\beta<1$, $\gamma>0$, and $N_0$ such that  
for every $N\geq N_0$, $\log^{(l)}N\leq k\leq N/2$, and $0<M\leq\exp(\gamma k)$, there exists a coloring $F:{[N]\choose k}\to [M]$ such that for every $V\sub[N]$, $|V|=2k$, $F$ restricted to ${V\choose k}$ is $\epsilon$-balanced, where $\epsilon=\beta^k$.
\end{corollary}


\bprf
Let $l,N$, $\log^{(l)}N\leq k\leq N/2$, $M\leq \exp(\delta l)$ be given.  
By Theorem~\ref{t-shift-extractor}, there exists an $F:\pw([N])\to [M]$ such that for every $V\sub[N]$, $|V|=2k$, the distribution of colors $F(X)$ for $X\in\pw([V])$ is $\alpha^k$-close to the uniform distribution on $[M]$ for some constant $\alpha<1$. We will show that the distribution of colors $F(X)$ for $X\in{[V]\choose k}$ is $\beta^l$-close to uniform distribution on $[M]$ for some $\beta<1$ that only depends on $\alpha$. 

Let $\vec{p}$ be the probability distribution on $[M]$ generated by $F$ from uniformly random $X\sub V$. The probability distribution  on $[M]$ generated by $F$ from uniformly random  $X\in{V\choose k}$ is $\vec{q}:=c\vec{p}\upharpoonright_{{V\choose k}}$ where $c=2^k/{2k\choose k}$. We have
\[
\|\vec{p}\upharpoonright_{{V\choose k}}-U_{\pw(V)}\upharpoonright_{{V\choose k}}\|_1
\leq\|\vec{p}-U_{\pw(V)}\|_1\leq \alpha^k.
\]
Hence $\|\vec{q}-U_{{V\choose l}}\|_1\leq c\alpha^k$. Since $c\approx\sqrt k$, we can bound $c\alpha^k\leq\beta^k$ with a suitable constant $\beta<1$ for $k$ sufficiently large. We can ensure that $k$ is sufficiently large by taking $N_0$ sufficiently large.
\eprf

For another Ramsey-theoretical result see Theorem~\ref{t-loss-less}.

\subsection{Bit-fixing sources and daisies}

The Ramsey-theoretical problems related to extractors for \emph{bit-fixing} sources are much less researched. These problems are stated in terms of \emph{daisies}.

\bdf
Let $Z$ be a set and $l<k\leq|Z|$ be numbers. An $(l,k)$-\emph{daisy} is a set of the form $\{U\cup X\ |\ X\in{V\choose l}\}$ for some $U,V\sub Z$ disjoint sets and $|V|=k$.
\edf
We define the \emph{Ramsey number for daisies} $D^l_M(k)$ to be the minimum number $N$ such that for every $F:\pw([N])\to[M]$, there exists a monochromatic $(l,k)$-daisy. 

Clearly, $D^l_M(k)\leq R^l_M(k)$. For $l=2$ and $M=2$, one can show an exponential lower bound using Lov\'asz Local Lemma. This is essentially all we know about $D^l_M(k)$. For all we know, $D^l_M(k)$ can be bounded by an exponential function in $k$ for all constants $l$ and $M$. For some results and open problems on daisies, see~\cite{???}.

The lack of methods to deal with daisies explains why we are not able to decide if there exist extractors for bit-fixing sources smaller than $\log\log N$.


\section{An extractor for zero-fixing sources of triply logarithmic size}\label{s-stepping}

In this section we present our construction based on the idea of the  \emph{stepping-up lemma} of Erd\H os and Hajnal~\cite{erdos-hajnal}. Given $k$ and $n$ they used binary trees to project $[2^{n+1}]$ on $[n]$ in such a way that from a coloring of ${[n]\choose k}$ without large monochromatic sets, one can construct a coloring of ${[2^{n+1}]\choose k+1}$ without large monochromatic subsets. We will use a similar projection mapping to reduce the construction of zero-fixing extractor on a set $A$ to a construction of a bit-fixing extractor on an exponentially smaller set $B$. Since a construction of extractors for $(n,k)$ bit-fixing sources are known for $k\approx \log\log n$, we obtain an extractor for $(n,k)$ zero-fixing sources with $k=O(\log\log\log n)$. To this end we show that the projection of a $(N,k)$-zero-fixing source is a convex combination of $(n,k')$-bit-fixing sources with a small error, where $k'=\Omega(k)$ and $N=2^{\Omega(n/k)}$.

We will prove the following:

\bt\label{t-steping-extractor}
{There exist constants $\delta_1,\delta_2>0$ and $C$ such that for every $N$ and $k$ such that $C\log\log\log N\leq k\leq \log N$
there exists an $\epsilon$-extractor $F:\{0,1\}^N\to\{0,1\}^m$ for $(N,k)$-zero-fixing sources where $m=\delta_1 k$ and $\epsilon=\max\{(\log N)^{-1},2^{-\delta_2k}\}$. }
The extractor is computable in polynomial time.
\et

\subsection{Trees}

Our main tool will be \emph{binary trees} with edges directed towards the root, which means that every node has indegree either 2 or 0. The 0-indegree verticas are \emph{leaves} (note that our leaves are \emph{vertices}, not edges). Furthermore, we will assume that the two children of each inner node are ordered. 
This induces a natural linear ordering on the leaves. In the rest of this section \emph{all trees are binary,} therefore we will often omit the specification ``binary''.

We will measure \emph{the size of a tree} by the number of its leaves; thus $|T|$ will denote the number of leaves of $T$.%
\footnote{This notation seems to be in conflict with our notation for the cardinality of sets, but notice that a binary tree with $k$ leaves can be represented by a set of $k$ binary strings.} 
The number of edges in a binary tree is $2|T|-2$.

Given a tree $T$ and a subset of leaves $X$, we will denote by $T_X$ the subtree of $T$ with leaves $X$ defined as follows. View $T$ as an ordered structure where the root is the the maximum and the leaves are minimal elements. This ordering defines an upper semilattice. Then $T_X$ is the  subsemilattice generated by $X$. We will call such subtrees \emph{leaf-generated subtrees}.

We will distinguish two types of leaves. A \emph{twin} is a leaf that shares a parent with another leaf (which in turn is also a twin). The other leaves will be called \emph{lone leaves.} A pair of twins sharing a parent will be called \emph{a twin pair}. There are at most $|T|-2$ lone leaves (and there are trees in which this bound is attained). Parents of lone leaves and twins will be called \emph{lone parents} and \emph{twin parents} respectively.

\bll{l-twins}
If $T$ is a tree and $X$ is a nonempty subset of leaves, then $T_X$ has at most as many twins as $T$.
\el
\bprf
By induction---if $T$ is not a single vertex, consider the two maximal proper subtrees of $T$.
\eprf
However, the number of lone leaves may increase.

The \emph{skeleton} of a tree $T$, denoted by $Sk(T)$, is the subtree leaf-generated by twins (see Figure~\ref{tree1}). The \emph{inner edges of $Sk(T)$}, the edges that are not connected to the leaves, will play a special role. The following is the key structural property of trees that we will use.

\bl
Every binary tree $T$ with at least two leaves can be represented as $Sk(T)$ extended with 
\ben
\item new nodes on the inner edges and leaves attached to them,
\item a chain with lone leaves attached on the root of $Sk(T)$.
\een
\el
\bprf
By induction---if $T$ has more than two leaves, consider the two maximal subtrees of~$T$.
\eprf

From Lemma~\ref{l-twins}, we have 
\[
|Sk(T_X)|\leq |Sk(T)|.
\]
 The number of inner edges of a skeleton is, clearly, $|Sk(T)|-2$, which is at most $|T|-2$ and when it is equal to $|T|-2$, then $T$ does not have any lone leaves.
Given a tree $T$ we will enumerate (starting with 1) the inner edges of $Sk(T)$ in a systematic way so that the edges in isomorphic skeletons are enumerated in the same way. We will denote the $i$-th inner edge of $Sk(T)$ by $e_i(T)$.

Let $T$ be a tree with leaves $L$ and let $\sigma:L\to\{0,1,*\}$. Then 
\[
T_\sigma:=T_{\{i|\sigma(i)\in\{*,1\}\}},
\]
i.e., $T_\sigma$ is the tree leaf-generated by leafs labeled by 1s and $*$s of $\sigma$. We will call the leaves of $T_\sigma$ labeled by $*$ \emph{free}.

\subsection{The projection mapping}

Suppose, w.l.o.g., that $k-1$ divides $n$. Let $T$ be the complete binary tree of depth $n/(k-1)+1$. Split the set $[n]$ into $k-1$ disjoint sets, say consecutive intervals, $D_0\dts D_{k-2}$ each of size $n/(k-1)$.
For $i=0\dts k-2$, let $\beta_i$ be a projection of the levels of $T$, excluding the level of leaves,\footnote{in fact, we also do not need the level next to the bottom one} 
onto $D_i$, i.e., for two non-leaf nodes $u,v\in T$ of different rank\footnote{the distance from the root}, 
$\beta_i(u)\neq \beta_i(v)$.

We will identify the set of leaves of $T$ with $[N]$, where $N=2^{n/(k-1)+1}$. Let $K\sub\{0,1\}^N$ denote the set of all vectors with at most $k$ ones. Alternatively, we can view $K$ as the set of characteristic vectors of subsets $X\sub[N]$ of size at most $k$.

The function $F_1$ maps $K$ on strings in $\{0,1\}^n$ with at most $k-2$ ones as follows. For $s\in K$,
\[
F_1(s):=b_0\cup b_1\cup\dots\cup b_j 
\]
where
\bi
\item $j$ is the number of inner edges of $Sk(T_s)$, 
\item $b_0=\{\beta_0(v_{0,1})\dts\beta_0(v_{0,l_0})\}$, where $v_{0,1}\dts v_{0,l_0}$ are the nodes of $T_s$ above the root of $Sk(T_s)$, and
\item for $i=1\dts j$, $b_i=\{\beta_i(v_{i,1})\dts\beta_i(v_{i,l})\}$, where $v_{i,1}\dts v_{i,l}$ are the nodes of $T_s$ on the edge $e_i(Sk(T_s))$.
\ei
(Any of $b_i$ may be empty; in fact all of them.) 
In plain words, we project the lone parents of $T_s$ to $[n]$, for each inner edge of $Sk(T)$, to a different part of $[n]$, and the lone parents of $T_s$ that are above the root of $Sk(T)$ to another part. Since the nodes on one $e_i(T)$ have different ranks, this ensures that the projection is bijective,%
\footnote{This is certainly not the most economical way to ensure bijectivity. E.g., we can omit $D_0$, because we can map the lone parents above the root of $Sk(T_s)$ to any block $D_i$, we can also omit $D_{k-2}$, because $b_{k-2}$ is always empty, etc.} see Figure~\ref{tree2}.

Let a $(N,k)$ zero fixing source defined by $\sigma$ be given. Let $V:=\{i\ |\ \sigma(i)=*\}$. The projections $F_1(s)$ for $s\in\{0,1\}^N$, $\sigma\sub s$,  do not form a zero-fixing source on $[n]$. The reason is that for different vectors $s$, the skeletons $Sk(T_s)$ may be different and thus the same lone parents may be mapped to different blocks $D_i$. Therefore we need to decompose the resulting source in such a way that on each part the skeleton is fixed while there are still enough parents of free leaves.

\subsection{The skeleton fixing procedure}

Let $T$ be a tree with leaves $L$, $|T|=k$. We will define a randomized procedure that produces a restriction $\rho:L\to\{0,1,*\}$ such that in $T_\rho$ all twins are fixed to 1. Our aim is to show that with probability close to 1 the the resulting tree $T_\rho$ has at least $\delta_1 k$ lone leaves for some $\delta_1>0$.

The procedure starts with $\rho=*^k$ and gradually extends $\rho$ by setting stars to zeros or ones. At each step the procedure checks if there is a twin in the restricted tree $T_\rho$ that still has a star. If there is no such twin, then it stops. If there is some, it picks a suitable one and sets it randomly to 0 or 1 with equal probability. We will specify the order in which twins are chosen when we prove the following lemma. When the procedure stops, all twins in $T_\rho$ are fixed to 1, which means that the skeleton is fixed. 

Note that we can view the resulting set of restrictions obtained as a binary decision tree; in particular, any two restrictions are incompatible.


\bll{l-skeleton-fixing}
There exist constants $\gamma<1$ and $\delta>0$ such that for every tree $T$, $|T|=k$, there exists a fixing procedure that with probability $\geq 1-\gamma^k$ produces a restriction $\rho$ such that all twins in $T_\rho$ are fixed to 1 and such that there are at least $\delta k$ lone leaves free (i.e. labeled by $*$s).
\el


We will first prove:
\bll{l-sfix}
Let $T$ be a tree of arbitrary size and let $\alpha$ be an assignment of $*$s and 1s to the leaves of $T$. Suppose $T$ has $\geq k/10$ lone leaves labeled with~$*$s. Then there is a fixing procedure as described above that starts with $\alpha$ and extends it to $\rho$ so that all twins in $T_\rho$ are fixed to 1s and with probability exponentially close to 1 there are at least $k/500$ lone leaves in $T_\rho$ still labeled $*$.
\el

\bprf
Let $A$ be the set of the parents of lone leaves of $T$ labeled $*$ by~$\alpha$. We consider two cases.
\ben
\item[{\bf (a)}] Suppose that there is an antichain $C\sub A$, $|C|\geq|A|/10$. Let $v\in C$ and let $l$ be its lone leaf and $S$ the neighbor tree of $l$, i.e., the two maximal proper subtrees below $v$ are $S$ and the single element tree $l$. In order for $l$ to be queried in the process, $l$ must become a twin, which means that $S$ must be reduced to a single node. There is at least one twin pair in $S$. 
The twins in this pair can be labeled by two 1s, one 1 and one $*$, or two $*$s. Thus the probability that both twins are fixed to 1s is at least $1/4$.
  Hence with probability at least $1/4$, $S$ will not be reduced to a single leaf and thus $l$ survives to the end (meaning that  $\rho(l)=*$ in the finial restriction). For two different nodes $v,u\in C$, the events that a twin pair is fixed are independent, because $v$ and $u$ are incomparable. Hence we can apply Chernoff's inequality and conclude that there are at least $|C|/5\geq k/500$ lone leaves $l$ with $\rho(l)=*$ in $T_\rho$ for the final restriction $\rho$ with probability exponentially close to 1.

\item[{\bf (b)}] Suppose that for every antichain $C\sub A$, $|C|<|A|/10$. Suppose that the procedure outputs $\rho$. For a node $v\in T$, we will denote by $\hat{v}$ its parent.

Let $D_1$ be the set of lone leaves $l$ of $T$ such that $\rho(l)=1$ and $\hat{l}\not\in T_\rho$. Hence if $l\in D_1$, then $l$ is the unique leaf below $\hat{l}$ that is fixed to 1 by $\rho$. This implies that $\{\hat{l}\ |\ l\in D_1\}$ is an antichain. By the assumption of this subcase, it follows that $|D_1|<|A|/10$. 

Let $D_2$ be the set of lone leaves $l$ of $T$ such that $\rho(l)=1$ and $\hat{l}\in T_\rho$. 

We claim that all leaves in $D_2$ are twins in $T_\rho$. Indeed, for $l$ to be fixed in the process it must first become a twin. That is, $l$ and some $v$ are twins in some $\sigma\sub\rho$. Since $\hat{l}\in T_\rho$, we also have $\hat{l}\in T_\sigma$. (Once a node disappears in the process, it is never restored.) Since $\hat{l}$ is the parent of $l$ in $T_\sigma$ it also is the parent of $v$. So $v$ must also be fixed to 1 in $\rho$, otherwise $\hat{l}$ would not be in $T_\rho$.

The parents of twins in $T_\rho$ are incomparable and since it is a subtree of $T$, they are also incomparable in $T$. This implies that $\{\hat{l}\ |\ l\in D_2\}$ is an antichain and $|D_2|<|A|/10$. 

Thus every $\rho$ fixes at most $\frac 2{10}$ of lone leaves. Since the process assigns zeros and ones randomly independently, with probability exponentially close to 1, it will not fix more than $\frac 12$ of the lone leaves of $T$. Hence  with probability exponentially close to~1, $T_\rho$ has at least $\frac 12 k/10$ lone leaves $l$ with $\rho(l)=*$.
\een

\eprf

\bprf (of Lemma~\ref{l-skeleton-fixing}).

\noindent {\bf Case 1.} $T$ has $\geq k/10$ lone leaves. Then we can apply Lemma~\ref{l-sfix} with $\alpha$ empty (all stars).

\medskip
\noindent {\bf Case 2.} $T$ has $< k/10$ lone leaves. Let $T'$ be the tree obtained from $Sk(T)$ by removing all twins of $T$. We consider two subcases.
\ben
\item[{\bf (a)}] Suppose that $T'$ has $\geq \frac 3{10}k$ lone leaves.
In this case the process will first query twins of $T$ that are attached to lone leaves of $T'$. When the first twin is queried, then it is fixed to 0 with probability $1/2$. If this happens, the second twin will become a lone leaf in the reduced tree. Thus we obtain with probability exponentially close to 1 at least $k/10$ free lone leaves (i.e., labeled $*$). Let $\alpha$ be the restriction that produces such a tree $T_\alpha$ with $k/10$ free lone leaves.  
In order to apply Lemma~\ref{l-sfix}, take $T_\alpha$ and $\alpha$ restricted to the leaves of $T_\alpha$. The restricted $\alpha$ only assigns 1s and $*$s and there are at least $k/10$ lone leaves with $*$s, so we can now apply Lemma~\ref{l-sfix}. The resulting fixing procedure is the composition of the procedure producing $\alpha$ and the procedure from Lemma~\ref{l-sfix}.

\item[{\bf (b)}] Finally, suppose that $T'$ has $< \frac 3{10}k$ lone leaves. Then there are $<\frac 6{10}k$ twins attached to the lone leaves of $T'$ (i.e., twins that together with lone leaves form subtrees with 3 leaves). Hence there are $\geq\frac 1{10}k$ twins of $T$ attached to twins of $T'$. Then there are at least $\geq\frac 1{40}k$ quadruples of twins attached to pairs twin pairs of $T'$. For each of these quadruples, we have probability $1/8$ that it will be fixed in such a way that one twin pair is fixed to ones and from the other one twin is fixed to 0 and the other remains free and becomes a free lone leaf (i.e., they will form a subtree with 3 leaves in which the twins are fixed to 1 and the lone leaf is $*$). Hence with probability exponentially close to 1, the resulting tree $T_\rho$ will have at least $\frac 1{160}k$ lone leaves.
\een
\eprf


\subsection{The extractor}

Let $X$ be a $(N,k)$-zero-fixing source given by a subset $V\sub [N]$, $|V|=k$. We will now describe the decomposition of $F_1(X)$ into a convex combination of bit-fixing sources on $[n]$. Each of the sources is a $k'$-bit-fixing source for some $k'\geq\delta k$, where $\delta>0$ is a constant, except for some sources whose total weight is exponentially small.

The source $X$ generates a random string $r\in\{0,1\}^V$ randomly uniformly. We can view the process of generating the random string $r$ as having two parts: first we run the skeleton fixing procedure to obtain some $\rho\in\{0,1,*\}^V$ and then we randomly extend it to a full vector $r\supseteq \rho$. The probability that we obtain $\rho$ is the weight of the source that $\rho$ produces (it is $2^{-t}$, where $t$ is the number of leaves set to 0 or 1 by $\rho$). Let $S$ be the set of lone leaves of $T_\rho$ and let $F_1(S)$ denote the projection of their parents to $[n]$. Then $F_1(s)\sub F_1(S)$ for every $s\supseteq\rho$, because the skeleton is fixed. Moreover, $F_1$ maps a 0-1 string defined on $S$ to a 0-1 string defined on $F_1(S)$ in a 1-1 way.%
\footnote{$F_1$ is defined on strings $K\sub\{0,1\}^N$, but now we focus on string of a given source where the strings are 0 outside of $V$, so we can view $F_1$ as defined on $\{0,1\}^V$.}
 Hence extensions of $\rho$ are mapped by $F_1$ to a $(n,k')$-bit-fixing source on $[n]$, where $k'$ is the number of stars in $\rho$. Note that it is a bit-fixing source, rather than zero-fixing one, because in $T_\rho$ there may be some lone leaves fixed to~1.

Now we are in a position to define our extractor and prove its properties. We use the extractor constructed by Cohen and Shinkar~\cite{cohen-shinkar}, see Theorem~5.1. in their paper. { They constructed an $\epsilon'$-extractor, which we will denote by $F_2:\{0,1\}^{n}\to\{0,1\}^m$, for $(n,k')$-bit-fixing sources which works for $k'\geq\log((\log n)/\epsilon'^2)+2\log\log((\log n)/\epsilon')+O(1)$.
Our extractor is the composition of $F_1$ with $F_2$ for $k'=\delta k$, where $\delta$ is the constant from the skeleton fixing procedure. Thus we get  an extractors for $(N,k)$-zero-fixing sources for $k=O(\log\log\log N)$. 
Furthermore, one can check that if $k'\leq\log N$ and $\epsilon\leq \max\{(\log N)^{-1},2^{-\delta_2k}\}$, then $F_2$ is computable in time sublinear in $N$. Since, clearly, $F_1$ is computable in polynomial time, $F:=F_2\circ F_1$ can also be computed in polynomial time.

To finish the proof of Theorem~\ref{t-steping-extractor}, it remains to compute the parameters $m$, $\epsilon$, and the time needed to compute the function $F$ in the whole range of parameters allowed in the theorem. 

We want to use an $\epsilon'$-extractor $F_2:\{0,1\}^{n}\to\{0,1\}^m$ for $(n,k')$-bit-fixing sources with the following parameters:
\ben
\item $k'=\log((\log n)/\epsilon'^2)+2\log\log((\log n)/\epsilon')+O(1),$
\item $m=k'-2\log(1/\epsilon')-O(1)$,
\item furthermore, $F_2(s)$ can be computed in time $n^{O(\log^2((\log n)/\epsilon'))}$.
\een

If we want to get error $\epsilon$ for our extractor, which
the composed function $F_2\circ F_1$, then we need to have the error $\epsilon'$ of $F_2$ slightly smaller, because part of the sources in the convex combination are not $(n,k')$-bit-fixing sources. The weight of the bad sources in the convex combination is exponentially small, so 
we have $\epsilon=\epsilon'+o(1)$. Note that even if $\epsilon$ were larger by a constant factor, the expressions above would still keep the same form if we replaced $\epsilon'$ by $\epsilon$, because the term $o(1)$ would be consumed by the big O.

Recall that we are projecting an $(N,k)$-zero-fixing source to $(n,k')$-bit-fixing sources. The construction gives us $n\leq k\log N$ and $k'=\Omega(k)$. Since we assume $k\leq\log N$ in the statement of the theorem, we have
\bel{e-n-N}
n\leq(\log N)^2.
\ee
We need to show three things:
\ben
\item[i.] $k$ can be as small as $O(\log\log\log N)$,
\item[ii.] $m=\Omega(k)$, 
\item[iii.] $F$ can be computed in polynomial time.
\een

To prove i., we will use 1. in the list of the properties of $F_2$. According to (\ref{e-n-N}), $\log\log n=\log\log\log N+O(1)$. We need to check that the contribution of the $1/\epsilon^2$ in 1. is also of the order $(\log n)^{O(1)}$. If $k=O(\log\log n)$ and $\epsilon\geq 2^{-\delta_2k}$, then we also have 
$k'=O(\log\log n)$ and $\epsilon'\geq 2^{-\Omega(k)}$, hence indeed, $1/\epsilon'^2=(\log n)^{O(1)}$.

To prove ii., it suffices to have  $\epsilon\geq 2^{-\delta_2k}$ for a sufficiently small $\delta_2$, because then the negative terms in 2. are smaller than $k'/2$.

Since $F_1$ is computable in polynomial time, in order to prove iii., we only need to bound the time for $F_2$. This amounts to substitute our bounds on $n$ and $\epsilon'$ into $n^{O(\log^2((\log n)/\epsilon'))}$. Let us first only estimate the expression without $\epsilon'$; we will use (\ref{e-n-N}).
\[
n^{O(\log^2(\log n))}\leq 2^{O(\log^2(\log((\log N)^2))\cdot(\log((\log N)^2)))}.
\]
This is, clearly, sublinear. The contribution of $\epsilon'$ will be 
\[
n^{O(\log^2(1/\epsilon'))}=2^{O(\log^2(1/\epsilon')\cdot(\log((\log N)^2)))}.
\]
Since in the theorem we assume $\epsilon\geq(\log N)^{-1}$, the resulting term is also sublinear.}
\qed


\section{Extractors based on shift graphs}\label{s-shift}

In this section we will present our second construction of extractors, based on colorings of shift graphs.

\bt\label{t-shift-extractor}
For every $l\in\N$, there exists $\alpha_l<1$, $\delta_l>0$ and $N_0$ such that for every $N\geq N_0$ and $\log^{(l)}N\leq k\leq N$, there exists an $\epsilon$-extractor $F:\{0,1\}^N\to\{0,1\}^m$ for $(N,k)$-zero-fixing sources, where 
$\epsilon=\alpha_l^k$ and $m=\lfloor\delta_lk\rfloor$.
Moreover, the extractor is computable in polynomial time if  $k\leq \beta\log\log N/\log\log\log N$ where $\beta>0$ is a constant.
\et 

\subsection{Shift graphs}

\bdf
Let $2\leq l\leq n-1$. The \emph{shift graph} $S=S(n,l)$ is a graph with  vertex set $V(S)={[n]\choose l}$ and edge set
\[
E(S)=\{\{\{x_1\dts x_l\},\{x_2\dts x_l,x_{l+1}\}\}|\ 1\leq x_1<x_2<\dots<x_{l+1}\leq n\}.
\]
\edf

The key property of shift graphs is that their chromatic numbers decrease exponentially with $l$. In order to express the upper bounds on the chromatic numbers we will use a function that is asymptotically equal to the binary logarithm. We define
\ben
\item $\blog x:=x$ for $x=1,2,3,4$ and
\item $\blog x:=m$ where $m$ is the integer satisfying  ${m-1\choose\lfloor\frac{m-1}2\rfloor}\leq x<{m\choose\lfloor\frac{m}2\rfloor}$ for $x\geq 4$.
\een
We note that $\blog x$ is nondecreasing and 
\bel{blog}
\blog x\approx\log_2 x.
\ee

Let $\chi$ denote the chromatic number of a graph. We will need the following facts.
\begin{fact}\label{f1}
If $\chi(S(n,l))\leq {m\choose\lfloor\frac{m}2\rfloor}$, then
$\chi(S(n,l+1))\leq m$.
\end{fact}
Since, trivially, $\chi(S(n,1))=n$, we have
\bel{coloring-shift}
\chi(S(n,l))=O(\log^{(l-1)}n).
\ee
\bprf(of Fact~\ref{f1})
Let $\psi:{[n]\choose l}\to{[m]\choose\lfloor m/2\rfloor}$ be a coloring, which means that $\psi(L_1)\neq\psi(L_2)$ whenever $(L_1,L_2)\in E(S(n,l))$. We define $\phi:{[n]\choose l+1}\to[m]$ as follows. For $1\leq x_1<x_2<\dots<x_{l+1}\leq n$, we choose 
$x\in\psi(x_1,x_2\dts x_l)\setminus\psi(x_2\dts x_l,x_{l+1})$, 
say the first such element, and set $\phi(x_1,x_2\dts x_l,x_{l+1}):=x$.

Now, if $((x_1\dts x_{l+1}),(x_2\dts x_{l+2}))\in E(S(n,l+1))$, then we have 
\[
\phi(x_1\dts x_{l+1})\in \psi(x_1\dts x_l)\setminus\psi(x_2\dts x_{l+1}),
\mbox{ and}
\]
\[
\phi(x_2\dts x_{l+2})\in \psi(x_2\dts x_{l+1})\setminus\psi(x_3\dts x_{l+2}).~~~~
\]
Consequently, $\phi(x_1\dts x_{l+1})\neq \phi(x_2\dts x_{l+2})$.
\eprf

\begin{fact}
If $\chi(S(n,l-1))\leq 4$, then $\chi(S(n,l+1))\leq 3$.
\end{fact}

\bprf(of Fact 2)
Consider a 4-coloring
\(
\psi:S(n,l-1)\to[4].
\)
We define $\phi:S(n,l+1)\to[3]$ as follows.  For $1\leq x_1<x_2<\dots<x_{l+1}\leq n$, set
\[\begin{array}{lll}
\phi(x_1\dts x_{l+1})&:=&\psi(x_2\dts x_l)\mbox{\quad if }\psi(x_2\dts x_l)\neq 4,\mbox{ otherwise}\\
\\
&:=&\mbox{some } j\in[4]\setminus\{\psi(x_1\dts x_{l-1}),\psi(x_2\dts x_l),\psi(x_3\dts x_{l+1})\}.
\end{array}\]
Consider $((x_1\dts x_{l+1}),(x_2\dts x_{l+2}))\in E(S(n,l+1))$. We distinguish two cases.
\ben
\item[a.] If $\psi(x_2\dts x_l)\neq 4$, then $\phi(x_1\dts x_{l+1})=\psi(x_2\dts x_l)$. On the other hand, $\phi(x_2\dts x_{l+2})$ equals either to $\psi(x_3\dts x_{l+1})$, or belongs to $[3]\setminus\psi(x_2\dts x_l)$. Since $\psi(x_2\dts x_l)\neq\psi(x_3\dts x_{l+1})$, we have in either case $\phi(x_1\dts x_{l+1})\neq\phi(x_2\dts x_{l+2})$.
\item[b.] If $\psi(x_2\dts x_l)= 4$, then $\psi(x_3\dts x_{l+1})\neq 4$ and we have
\[\begin{array}{lll}
\phi(x_1\dts x_{l+1})&\in&[3]\setminus\{\psi(x_1\dts x_{l-1}),\psi(x_3\dts x_{l+1})\}\\
\phi(x_2\dts x_{l+2})&=&\psi(x_3\dts x_{l+1}).
\end{array}\]
Consequently $\phi(x_1\dts x_{l+1})\neq\phi(x_2\dts x_{l+2})$.
\een
\eprf

It follows from these facts that for every $n$ there exists an integer $l$ such that $\chi(S(n,l)))\leq~3$. On the other hand we have:
\begin{fact}\label{fact3}
If $n\geq 2l+1$, then $S(n,l)$ contains an odd cycle and consequently $\chi(S(n,l))\geq 3$.
\end{fact}
\bprf(of Fact 3)
Let $n\geq 2l+1$. Then the sets
\[\begin{array}{l}
\{1,2\dts l\},\{2\dts l,l+1\}\dts \{l+1\dts 2l\},\{l+2\dts 2l+1\},\\
\\
\{l,l+2\dts 2l\},\{l-1,l,l+2\dts 2l-1\}\dts\{2,3\dts l-1,l,l+2\}
\end{array}\]
form an odd cycle in $S(n,l)$.
\eprf

The bound from Fact~\ref{f1} was first proved for infinite cardinals be Erd\H{o}s and Hajnal~\cite{erdos-hajnal}. The version for finite cardinals, the one above, appeared in~\cite{harner-entringer}. Fact 2 has not appeared in the literature, but a similar idea was used by Schmerl~\cite{schmerl},  Poljak~\cite{poljak}, and  Duffus, Lefman, and R\"odl~\cite{duffus-lefmann-rodl}.  


\subsection{Special symbol-fixing sources and their extractors}

A symbol-fixing source, introduced in~\cite{kamp-zuckerman}, is like a bit-fixing source except that the alphabet of the strings is larger than 2. We will introduce an auxiliary concept that we need in our construction, which is a sort of cross-bread between a symbol-fixing source and a bit-fixing source.

\bdf
A  \emph{special $(n,k,d)$-symbol-fixing source} $X$ is a random variable producing strings from  $[d]^n$ of the following form. For some string $\sigma\in( [d]\cup{[d]\choose 2})^n$ that has exactly $k$ pairs from ${[d]\choose 2}$, $X$ produces strings $s\in [d]^n$ that are consistent with $\sigma$ each with probability $2^{-k}$. We say that $s$ is consistent with $\sigma$ if $s_i=\sigma_i$ or $s_i\in\sigma_i$ for every $i=1\dts n$.
\edf
In plain words, values on some coordinates are fixed and on other coordinates there are two values allowed; the two values may be different for different coordinates.

\medskip
\noindent{\bf Example.} Let $d=3,n=3$ and $\sigma=(1,\{1,2\},\{2,3\})$. Then the source produces strings $112,113,122,123$ with equal probability.

\bll{l-symbol-fixing}
For every $\epsilon,n,k,m$ and $d$, if $1<m\leq k\leq n$ and 
\[
d\leq\exp_2\left(\frac{\frac{\log{\rm e}}3\cdot \epsilon^2\cdot 2^k-2^m-1}{2n}\right),
\]
then there exists an $\epsilon$-extractor $F$ for  special $(n,k,d)$-symbol-fixing sources with $m$ outputs bits, i.e., $F:[d]^n\to\{0,1\}^m$.
\el
\bprf
This lemma is proved by a standard counting argument. We need to recall an equivalent definition of the total variation distance.
\[
{\rm d}(\mu,\nu)=\sup_A|\prob_\mu(A)-\prob_\nu(A)|.
\]
The supremum is over all events $A$.%
\footnote{In this paper the supremum is always the maximum, since we only consider finite probability spaces.}

Let $X$ be a special $(n,k,d)$-symbol-fixing source. The random variable $X$ produces strings from some set $S$ of size $2^k$, each string with the same probability $2^{-k}$ (this is all we need to know about $X$).
Let $Y$ be distributed uniformly on $\{0,1\}^m$. Let $A\sub\{0,1\}^m$ be an arbitrary event on $\{0,1\}^m$. Consider random function $F:[d]^n\to\{0,1\}^m$. 
We need to bound the following probability
\[
\prob_F[|\prob[A(F(X))]-\prob[A(Y)]|>\epsilon].
\]
The outer probability is, as indicated, with respect to randomly chosen function $F$. The term $\prob[A(F(X))]$ is the number of strings $s$ from $S$ such that $F(s)\in A$ divided by $|S|$, which is $2^k$. The term $\prob[A(Y)]$ is the probability that a random string $t$ chosen from $\{0,1\}^m$ is in $A$, which is $|A|/2^m$; let us denote it by $p$. Since $S$ is fixed, we only need to know the values of $F$ on this set. For a given $s\in S$, we have $\prob_F[A(F(s))]=p$ and for $s,s'\in S$, $s\neq s'$, the events $A(F(s))$ and $A(F(s'))$ are independent. Hence we can apply the Chernoff bound, which gives us 
\[
\prob_F[|\prob[A(F(X))]-\prob[A(Y)]|>\epsilon]\leq 2{\rm e}^{-\frac{\epsilon^22^k}3}.
\]

We will use this bound to show that there exists an $F$ such that $|\prob[A(F(X))]-\prob[A(Y)]|\leq\epsilon$ for \emph{every} $(n,k,d)$-symbol-fixing source and \emph{every} event $A$. This property of $F$ is equivalent to being an $\epsilon$-extractor for such sources.

The number of events $A$ is $2^{2^m}$. Let $K$ denote the number of  special $(n,k,d)$-symbol-fixing sources. Then, by the union bound, the probability that $|\prob[A(F(X))]-\prob[A(Y)]|>\epsilon$ for some source $X$ and some predicate $A$ is bounded by
\[
 2{\rm e}^{\frac{-\epsilon^22^k}3}\cdot 2^{2^m}\cdot K.
\]
The number of special $(n,k,d)$-symbol-fixing sources can be bounded by
\[
K\leq \left(d+{d\choose 2}\right)^n\leq(d^2)^n=d^{2n}.
\]
Hence, there exists an $\epsilon$-extractor if
\[
2{\rm e}^{\frac{-\epsilon^22^k}3}\cdot 2^{2^m}\cdot d^{2n}<1.
\]
\eprf

\begin{corollary}\label{cor5.3}
For every $c_1$, there exists $\alpha<1$, $\delta>0$, and $c_2$ such that for every $m,n,k,d$ such that  $1<m\leq\delta k$, $c_2\log n\leq k\leq n$, $d\leq c_1n$, there exists an $\epsilon$-extractor for $(n,k,d)$-special symbol fixing sources with $m$ output bits and $\epsilon=\alpha^k$.
\end{corollary}


\subsection{The extractor}

The extractor will again be constructed as a composition of two functions $F_1$ and $F_2$. The first function transforms an $(N,k)$-zero-fixing source into a special symbol-fixing source, the second one is an extractor for symbol-fixing sources. The essential difference is that now the size of domain of $F_2$ only depends on $k$ and it is not much larger than $k$.

Let $l\geq 2$ be a constant, let $k$ and $N$ be such that 
$\log^{(l-1)}N\leq k< \log^{(l-2)}N$, and suppose $N$ is sufficiently large. Let $\psi$ be a coloring of the shift graph $S(N,l)$ by $d$ colors where $d=O(k)$. Such a coloring exists by (\ref{coloring-shift}), since $k\geq \log^{(l-1)}N$. Let $p:=\lfloor\frac {k-1}l\rfloor$. Define a mapping 
\[
F_1:{[N]\choose\leq k}\to[d]^p,
\]
by putting, for $X\sub[N]$, $|X|\leq k$,
\[
F_1(X)=(\psi(X_1),\psi(X_2)\dts\psi(X_j),1\dts 1),
\]
where
\[
\begin{array}{l}
X=X_1\cup\dots\cup X_j\cup Z,\\
\\
|X_1|=\dots |X_j|=l,\ |Z|<l,\\
\\
X_1<X_2<\dots<X_j<Z.
\end{array}
\]

\bl
Let $k':=2^{-2l-3}p$. If $X$ is a $(N,k)$-zero-fixing source, then $F_1(X)$ is a convex combination of  special $(p,t,d)$-symbol-fixing sources where the total weight of sources with $t<k'$ is exponentially small, $2^{-\Omega(k)}$.
\el
\bprf
Let a  $(N,k)$-zero-fixing source be given by some $V\sub[N]$, $|V|=k$. Let 
\[
V=I_1\cup I_2\cup\dots\cup I_q,\quad I_1<I_2<\dots <I_q,
\]
be a partition of $V$ into blocks of sizes $|I_{2i+1}|=l+1$, $|I_{2i}|=l-1$, with the exception that the last block $I_q$ may be smaller. According to our choice of $p$, the number of blocks with odd indices and size $l+1$ is 
$\lceil p/2\rceil$.

Let $X$ be a random subset of $V$ (generated by our zero-fixing source). For odd $i$, let $A_i(X)$ be the event defined by the conjunction of the following three clauses:
\[\begin{array}{lr}
|X\cap(I_1\cup\dots\cup I_{i-1})|\equiv 0\mod l,&\qquad\mbox{(C1)}\\ \\
X\cap I_{i-1}\mbox{ is an initial segment of }I_{i-1},&\qquad\mbox{(C2)}\\ \\
X\cap I_i=I_i\setminus\{\max I_i\}\mbox{\quad or\quad }X\cap I_i=I_i\setminus\{\min I_i\}.&\qquad\mbox{(C3)}

\end{array}
\]
For $i=1$, clauses C1 and C2 are always true, hence $\prob[A_1(X)]=2^{-l}$, and for every odd $i\geq 3$ and $Y\sub I_1\cup\dots\cup I_{i-2}$,
\bel{prI}
\prob[A_i(X)\ |\ X\cap(I_1\cup\dots\cup I_{i-2})=Y]= 2^{-2l-1},
\ee
because this probability is equal to 
\[
\prob[X\cap I_{i-1}=Z\mbox{\ and\ C3}\ |\ X\cap(I_1\cup\dots\cup I_{i-2})=Y],
\]
where $Z$ is the initial segment of $I_{i-1}$ such that $|Y\cup Z|\equiv 0\mod l$.
Hence ${\rm E}[|\{i\ |\ A_i(X)\}|]\geq 2^{-2l-2}p$. Furthermore, since the probability in~(\ref{prI}) is $2^{-2l-1}$ independently of $Y$,%
\footnote{This is the reason why we have clause (C2). Without this clause the argument would be more complicated, because we would not be able to use the Chernoff inequality, although we might get a better constant by a more complicated argument.}
 the events $A_i(X)$, for $i$ odd, are independent. Thus we have, by the Chernoff inequality and recalling that $k'=2^{-2l-3}p$,
\bel{prA}
\prob[|\{i\ |\ A_i(X)\}|< k']\leq {\rm e}^{-p/8}.
\ee

We will now define a decomposition of $F_1(X)$ into a convex combination of special $(p,t,d)$ symbol-fixing sources. A source in this combination is given by
\ben
\item a subset $J$ of odd integers in $[p]$ and
\item sets $Y_i\sub I_i$ for $i\in[p]\setminus J$
\een
such that 
\bi
\item for $i\in J$, $|\bigcup_{j<i,j\not\in J}Y_j|\equiv 0\mod l$ and $Y_{i-1}$ is an initial segment of $I_{i-1}$ if $i\geq 3$,
\item for $i$ odd $i\not\in J$, $|\bigcup_{j<i,j\not\in J}Y_j|\not\equiv 0\mod l$, or $Y_i\not\in\{I_i\setminus\{\max I_i\},I_i\setminus\{\min I_i\}\}$.
\ei
The source determined by $(J,\{Y_i\}_{i\not\in J})$ produces uniformly independently all $X\sub V$ such that
\ben
\item for all $i\not\in J$, $X\cap I_i=Y_i$, and
\item for all $i\in J$, either $X\cap I_i=I_i\setminus\{\max I_i\}$ or $X\cap I_i=I_i\setminus\{\min I_i\}$.
\een
Let $X$ be produced by this source, i.e., $X$ satisfies 1. and 2. above. Let $X=X_1\cup\dots\cup X_j\cup Z$ be the partition of $X$ into segments of length $l$, except for $Z$. Then for every $i\in J$, there is an $i'$ such that   $X_{i'}=I_i\setminus\{\max I_i\}$ or $X_{i'}=I_i\setminus\{\min I_i\}$. Hence $\psi(X_{i'})$ is either $\psi(I_i\setminus\{\max I_i\})$ or $\psi(I_i\setminus\{\min I_i\})$ and these two colors are different. The blocks $X_{i'}$ that are not associated with any $I_i$ in this way are fixed. Hence $(J,\{Y_i\}_{i\not\in J})$ determines a special  $(p,t,d)$ symbol-fixing source.

Note that the weight of the source is the probability that a random $X$ satisfies~1. and~2. The probability in the inequality~(\ref{prA}) is the probability that a random $X$ satisfies these conditions for some source $(J,\{Y_i\}_{i\not\in J})$ with $|J|<k'$. Thus we have shown that the total weight of the special $(p,t,d)$ with $t<k'$ is exponentially small.

\eprf

To finish the proof of Theorem~\ref{t-shift-extractor}, we only need to compose $F_1$ with an extractor $F_2$ for special $(p,k',d)$ symbol-fixing sources whose existence follows from Corollary~\ref{cor5.3} where we take
\[\baa{l}
n:= p= \lfloor\frac{k-1}l\rfloor,\\
m:= \lfloor\delta k'\rfloor,\\
k:= k'=p 2^{-2l-3},\\
d:= O(k)=O(k').
\ea\]


The resulting $F$ is not explicitly defined, because we do not have an explicit definition of $F_2$. However, since $F_1$ is computable in polynomial time and a brute force search for $F_2$ can be done in polynomial time if $k$ is sufficiently small, we obtain a polynomial time algorithm for $F$ for all  sufficiently small $k$. We will now estimate how small $k$ should be. We have to search through all functions $F_2:[d]^p\to 2^m$. Here we have $d=O(k)$, $p,m<k$. Hence the number of such functions is $\leq 2^{2^{O(k\log k)}}$. Since the time needed to test each function is negligible w.r.t. the number of functions, the total time can also be bounded by $\leq 2^{2^{O(k\log k)}}$. Thus there exists $\beta>0$ such that the time needed for the search is polynomial if $k\leq \beta\log\log n/\log\log\log n$.
\qed


\subsection{A loss-less disperser}

The following version of our construction can produce only $o(k)$ bits of entropy, but it has the interesting feature that it is a \emph{loss-less disperser}, by which we mean that all possible values are always present. Although it also holds true for colorings of ${[n]\choose\leq k}$ it is more natural to state it for $k$-tuples.

Let $\lambda(n)$ be the minimal $l$ with $\chi(S(n,l))\leq 3$. It follows from~(\ref{blog}) that
\[
\lambda(n)=(1+o(1))\log^*n.
\]

\btl{t-loss-less}
Let $\lambda(n)\leq k\leq n$. Then there exists an efficiently computable function $F:{[n]\choose{k}}\to [m]$, where $m= 3^{\lfloor k/\lambda(n)\rfloor}$, such that for every $V\sub[n]$, $|V|=2k+\lfloor k/\lambda(n)\rfloor$, $F$~maps ${[V]\choose k}$ onto $[m]$.
\et
\bprf
We will use essentially the same mapping as $F_1$ in the construction of our extractor, except that we now take $l$ large enough for the shift graph to be colorable by three colors. In more detail, let $l:=\lambda(n)$ and assume w.l.o.g. that $l$ divides $k$. Let $X\in{[n]\choose k}$. Divide $X$ into consecutive parts $X_1\dts X_{k/l}$ of size $l$ and define
\[
F(X):=(\gamma(X_1)\dts\gamma(X_{k/l})),
\]
where $\gamma$ is the three-coloring of $S(n,l)$.

Let $V\sub[n]$,  $|V|=2k+k/l$ be given. Divide $V$ into $k/l$ consecutive parts $V_1\dts V_{k/l}$ of size $2l+1$. On each block $V_i$, each of the three colors must appear for some $Y\sub V_i$, $|Y|=l$, by Fact~\ref{fact3}. Hence for every vector $v\in [3]^{k/l}$ we can pick sets $X_1\sub V_1\dts X_{k/l}\sub V_{k/l}$ such that
$F(X_1\cup\dots\cup X_{k/l})=v$.
\eprf

\section{Upper bounds on the available entropy}\label{s-upper-bounds}

In this section we will prove that if $N$ is slightly more than $i$-times iterated exponential, then for every $F:\{0,1\}^N\to \{0,1\}^k$ there exists a $(N,k)$-zero-fixing source $X$ such that $F(X)$ has at most $k-i+O(i/2^{k-i})$ bits of entropy.

\bigskip
For a finite nonempty set of integers $X$, we denote by $\partial X:=X\setminus\{\max X\}$.

\bll{l-derivative}
Let $k,n,m,N$ be such that $k\leq n$ and  $N\geq n\cdot m^{{n-1\choose \leq k-1}}$.%
\footnote{${n-1\choose \leq k-1}$ denotes $\sum_{i\leq k-1}{n-1\choose i}$.}
 Then for every $\varphi:{[N]\choose \leq k}\to[m]$ there exists $V\sub[N]$, $|V|=n$, such that for every $X\sub V$, $X\neq\emptyset$, $\varphi(X)$ depends only on $\partial X$.
\el
The latter condition means that  $\varphi(X)=\varphi'(\partial X)$ for some function $\varphi':{[N]\choose \leq k-1}\to[m]$.

\bprf
Let $k,n,m,N$ and $\varphi$ satisfying the assumption be given. We will describe the construction of $V$ by the following pseudocode.

\bigskip


\fbox{%
\begin{minipage}{5 in}
\begin{tabbing}
1. $V:=\emptyset$, $U:=[N]$\\
2. $c:=$ the most frequent $c=\varphi(\{u\})$ for $u\in U$\\
3. $U:=\{u\in U\ |\ \varphi(\{u\})=c\}$\\
4. $V:=\{\min U\}$, $U:=U\setminus\{\min U\}$\\
5. do while \= $|V|<n$ and $U\neq\emptyset$:\\
6.         \> do for all \= $X\sub V$ such that $1\leq |X|\leq k$ and $\max X=\max V$:\\
7.         \>            \> $c:=$ the 
                           most frequent $c=\varphi(X\cup\{u\})$ 
                           for $u\in U$\\
8.         \>            \> $U:=\{u\in U\ |\ \varphi(X\cup\{u\})=c\}$\\
9.         \> $V:=V\cup\{\min U\}$, $U:=U\setminus\{\min U\}$\\
10. output $V$
\end{tabbing}
\end{minipage}}


\bigskip
It is clear that the algorithm produces a set $V$ with the required properties if the loop reaches some $V$ such that $|V|=n-1$ while $U$ is still nonempty. So we only need to estimate how big $N$ suffices. Since we have $m$ colors, the size of $U$ at 3. is at least $N/m$. Then at 4. it decreases by one. Similarly in the loop 6., the size of $U$ decreases at most by a factor $m^{{|U|-1\choose \leq k}-1}$ and then at 9. it decreases by one. Each division (in 3. and 8.) can be coupled with a subset $Y$ of $V\setminus \{\max V\}$, $|Y|\leq k-1$, where $V$ is the output $V$. Similarly, each subtraction of 1 is coupled with an element of $V\setminus \{\max V\}$. Hence we can lower bound the size of $U$ at the end of the procedure (when 9. is reached for the last time) by a number obtained from $N$ by ${n-1\choose\leq k-1}$ divisions by $m$ interleaved by $n-1$ subtractions of 1. If we postpone subtracting 1 to a later stage, we, clearly, get a smaller (or equal) number. Hence, for the lower bound, we can assume that all subtractions are done at the end. Thus in order for the algorithm to produce a $V$ with properties required, it suffices that
\[
N/m^{n-1\choose\leq k-1}-(n-1)\geq 1,
\]
which gives us the bound stated in the lemma.
\eprf

\bll{l-x}
Let $k,m,i,N$ be numbers such that $i\leq k$ and 
\[
N\geq m^{\exp^{i-1}_{m^k}({2^{k-i+1}})}.   
\]
Then for every $\varphi:{[N]\choose \leq k}\to[m]$ there exists a $V\sub[N]$, $|V|=k$ such that 
\ben
\item for subsets $X\sub V$ of cardinality $\leq k-i$ their color $\varphi(X)$ only depends on their cardinality (i.e., $\varphi(X)=\alpha(|X|)$ for some function $\alpha:\N\to[m]$),
\item for subsets $X\sub V$ of cardinality $> k-i$ their color $\varphi(X)$ does not depend on the last $i$ elements of $X$ (i.e., $\varphi(X)=\varphi^{(i)}(\partial^iX)$ for some function $\varphi^{(i)}:{\cal P}(\partial^iV)\to[m]$).
\een
\el
\bprf
This lemma follows by repeated applications of Lemma~\ref{l-derivative}. Namely, we first obtain $\varphi'$ from $\varphi$ and all one-element sets have the same $\varphi$-color. Then we apply the lemma to $\varphi'$; we get $\varphi''$ and all one element sets get the same $\varphi'$-color, hence all two-element sets get the same $\varphi$-color; and so on.

So it remains to estimate how big $N$ suffices for performing these operations.
To this end we need to simplify the bound from Lemma~\ref{l-derivative}. We will use two bounds:
\ben
\item $n\cdot m^{{n-1\choose \leq k-1}}\leq m^{2^{n-1}+\log n/\log m}\leq m^{2^n}$,
\item $n\cdot m^{{n-1\choose \leq k-1}}\leq m^{(n-1)^{(k-1)}+\log n/\log m}\leq m^{n^k}$,
\een 
for $m,n\geq 2$ and $k\geq 1$.

In the last step we need $V$ of size $n=k-i+1$. So using 1., it suffices to take $N_1=m^{2^{k-i+1}}$. Assuming we have shown that in the $j$th step before the end it suffices to have
\[
N_j=m^{\exp^{j-1}_{m^k}({2^{k-i+1}})},
\]
then according to 2., it suffices to put
\[
N_{j+1}=m^{N_j^k}=m^{\left(m^{\exp^{j-1}_{m^k}\left({2^{k-i+1}}\right)}\right)^k}
=m^{m^{k\cdot\exp^{j-1}_{m^k}\left({2^{k-i+1}}\right)}}=m^{(m^k)^{\exp^{j-1}_{m^k}\left({2^{k-i+1}}\right)}}
=m^{\exp^{j}_{m^k}\left({2^{k-i+1}}\right)}
\]

\eprf

The following bound can easily be proven by induction: for all $i\geq 0$, $x\geq 1$ and $r\geq 2$,
\bel{e-exp}
\exp^i_r(x)\leq\exp^i_2(x\log r+\log\log r+1).
\ee

\bt
Let $k,m,i,N$ be numbers such that $k\geq 2$, $i\leq k$, $2\leq m\leq 2^k$ and 
\[
N\geq \exp^{i+1}_2(k+2\log k+2).
\]
Then for every $\varphi:{N\choose \leq k}\to[m]$ there exists a $V\sub[N]$, $|V|=k$ such that the number of colors of $\varphi(X)$ for subsets $X\sub V$ is at most $2^{k-i}+i$. Hence the entropy of $\varphi(X)$ on such a source $X$ is at most $\log(2^{k-i}+i)$.
\et
\bprf
The theorem follows from the previous lemma by observing that if $X\sub V$ then $\partial^i X\sub\partial^iV$ for subsets $X$ with at least $i$ elements and $|\partial^iV|=k-i$. Hence these sets have at most $2^{k-i}$ $\varphi$-colors. The sets with $<i$ elements have at most $i$ colors, because their colors only depend on their cardinalities.

It remains to show that the expression in Lemma~\ref{l-x} can be bounded by the one in the theorem, where $m\leq 2^k$. Using $m\leq m^k$ and the inequality~(\ref{e-exp}), we can bound it by
\[
\leq\exp^i_{m^k}(2^{k-i+1})\leq \exp^i_2(2^{k-i+1}(\log m^k+\log\log m^k+1))
\leq \exp^i_2(2^{k-i+1}(k^2+2\log k+1))
\]
\[
\leq \exp^{i+1}_2(k+2\log k+2).
\]

\eprf


\section{Conclusions and open problems}

For $k$ being a finite number iterated logarithm of $n$, our extractors extract a positive fraction of entropy from $(n,k)$-zero-fixing sources. On the other hand the upper bounds on the amount of entropy that can be extracted only show that with each logarithm there is a loss of approximately one bit of entropy. Can one narrow down this gap? In this paper we have not tried hard to make the fraction of extracted entropy as large as possible. One can certainly get larger fractions of the available entropy by analyzing our constructions more carefully, but we do not see how one can get the amount of extracted entropy close to $k$, say $0.9k$. We think that new ideas are needed to this end.

The biggest challenge is to construct extractors for small \emph{bit-fixing} sources. We hope that our constructions will eventually help construct also extractors for small bit-fixing sources, but it is also possible that it will require developing completely new methods. If new methods are needed, the natural starting point is studying daisies.

Another possible research project is to extend our first construction to smaller zero-fixing sources. What prevents us from iterating the stepping up lemma is that we need a bit-fixing extractor to which the stepping-up construction is applied. But note that we only need to fix a small number of bits to~1. So it is possible that our construction can be adapted to construct extractors for bit-fixing sources with small number of bits fixed to~1 and then we would be able to iterate the stepping-up process.%
\footnote{We know that this is possible if instead of extractors one only requires that the functions produce positive fractions of available entropy.} 
This may be the first step towards a construction of extractors for bit-fixing sources with arbitrary number of bits fixed to~1.



\psfig{A binary tree with its skeleton consisting of black nodes}{tree1}{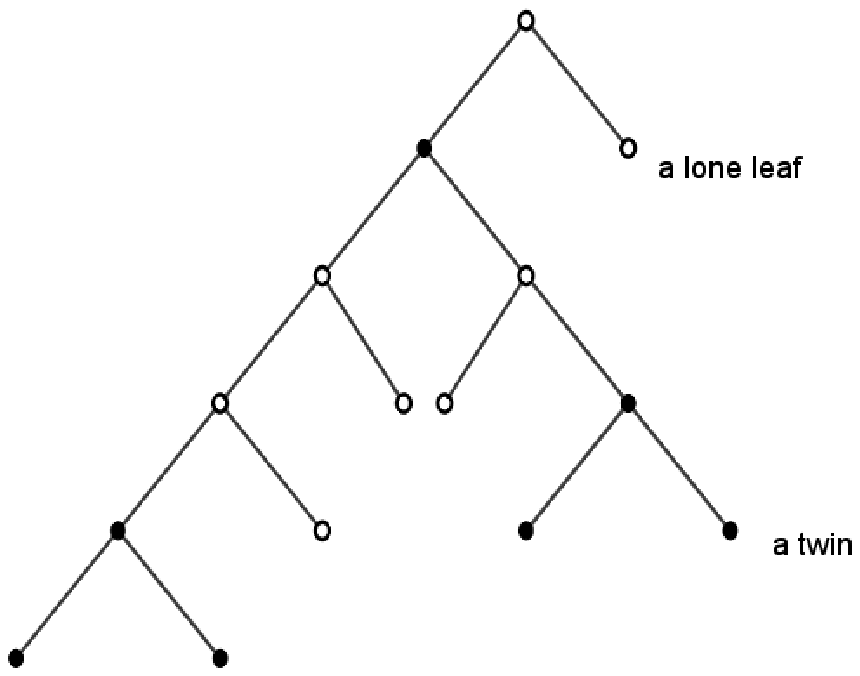}

\psfig{The projection mapping}{tree2}{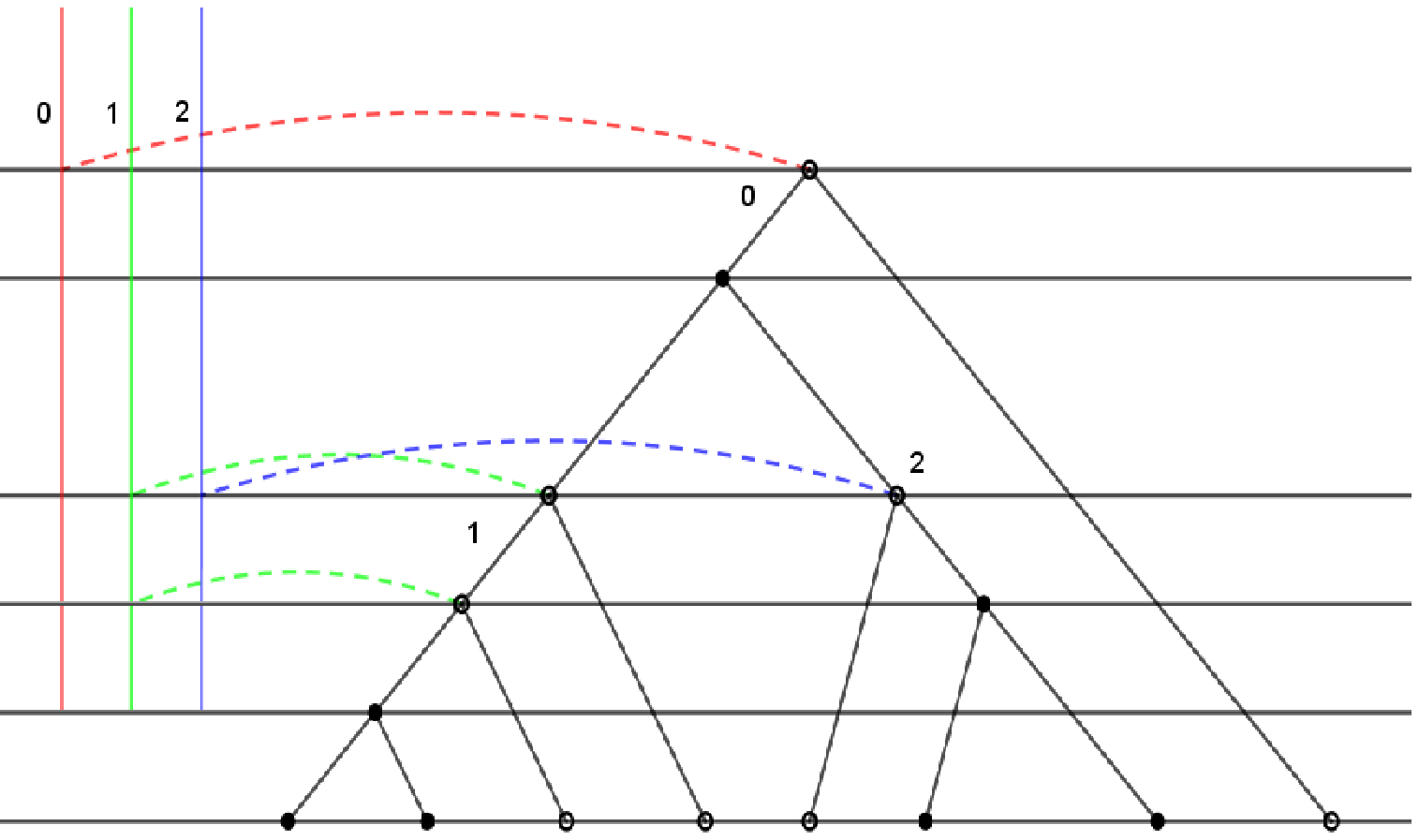}

\end{document}